\documentstyle[prl,preprint,aps]{revtex}
\begin{document}
\draft
\title{3D-2D phase transition in vortex lattice of layered
high-temperature superconductors of BSCCO
(Bi$_{1.7}$Pb$_{0.3}$Sr$_{2}$Ca$_{2}$Cu$_{3}$O$_{4}$)}
\author{J. G.  Chigvinadze,  A. A. Iashvili, T. V. Machaidze}
\address{E. Andronikashvili Institute of Physics, 380077 Tbilisi, Georgia}
\date{\today}
\maketitle
\begin{abstract}
Using the high-sensitive mechanical method of measuring
\cite{jab73} the dissipation processes in superconductors, the
3D-2D phase transition in magnetic field was investigated in the
vortex lattice of strongly anisotropic high-temperature
superconductors of Bi(2223) system. It was shown that with the
increase of external magnetic field the dissipation connected
with a motion of vortexes relative to the crystal lattice of a
sample, first increases, reaching the maximum and then with a
further increase of the magnetic field sharply reduces as a result
of pinning enhancement in connection with the transition of
three-dimensional vortex  system in quasi-two-dimensional
"pancake" one.
\end{abstract}

With a help of computer calculations it was shown recently in
\cite{ols00} that in the layered high-temperature superconductors
of BCCO system the disintegration of three-dimensional vortexes into
"pancake" ones  - the so-called 3D-2D transition could take place.

This transition must be accompanied by a sharp increase of
critical current that in its turn means a sharp increase of a
pinning force accompanying the disintegration of volume vortexes
into "pancake" - like ones. This also should strongly affect the
dissipation processes accompanying the motion of vortex system.

Naturally, at this transition, as a result of the increase of the
pinning force, the energy of dissipation should also sharply
diminish.

If one measures the losses in a vortex system their strong
decrease would be obtained at the 3D-2D transition.

The experiment on investigations of such type phenomena was made
by us using the oriented layered high-temperature superconductor
samples of $Bi_{1.7}Pb_{0.3}Sr_{2}Ca_{2}Cu_{3}O_{y}$
[BSCCO(Bi2223)] system.

Layered high-temperature superconductors are characterized by a
strong anisotropy of their superconductive properties. From this
point of view the layered high-temperature superconductors of BCCO
system are particularly distinguished. They are characterized by a
high degree of anisotropy \cite{feig90,vin90,far89} with anisotropy parameter
$\Gamma$=M/m$\gg$1, (where M is the effective mass along c-axis, and m is
the effective mass in the ab-plane) and amounts to $\Gamma\approx10^{3}$.

When outer field H is directed along c-axis, {\bf H}$\parallel${\bf c}, then below a
definite field $B=B_{2D}$ the vortex system behaves as
three-dimensional one, and above ${\rm B>B_{2d}}$ it behaves like
quasi-two-dimensional system.

The 3D-2D transition field $B=B_{2D}$ can be evaluated by expression \cite{feig90}:
\begin{equation}
{\rm B}_{2D}  \approx \frac{{\Phi _0 }}{{\Gamma d^2 }}
\end{equation}
Where d is the distance between CuO layers, and $\Phi_{0}$ is the
magnetic flux quantum - $\Phi_{0}=2.07\cdot10^{-7}$ G$\cdot
cm^{2}$. The 3D-2D transition from a three dimensional character
of superconductor behavior to the quasi-two-dimensional one takes
place when
\[
\xi _ \bot   = d \,\, ,
\]
where $\xi_{\bot}$ is the coherence length perpendicular to the
{\bf ab}-plane, that is the coherence length along {\bf c}-axis
$$
\xi _ \bot  =  \xi _c \,\, ,
$$
and d is the distance between layers (d=c/2). From this condition
we could define the so-called crossover temperature - T$_{cr}$
\cite{far89} which proves to be equal:
\begin{equation}
T_{cr}   =  \left[ {1 - \frac{{2\xi _{c^{} } ^2 (0)}}{{d^2 }}}
\right]T_c
\end{equation}
 T$_{cr}$ is always below the critical temperature of superconductive
transition  T$_{c}$ ( T$_{cr} < T_{c}$).

When $\xi_{c}(T) \ll d$ one could apply the
quasi-two-dimensional approximation and the 3D-2D transition would
be observed below $T<T_{cr}$, i.e. when
$T_{c}(T_{cr}<T<T_{c})$.

Consequently, the processes taking place under these conditions are
described by the layered model. And if the temperature T$> T_{cr}$,
but below $T_{c}(T_{cr}<T<T_{c})$, then properties of
superconductors can be described by the unequilibrium
anisotropic 3D model.

As the estimation shows, in Bi (2223) system the
3D-2D transition can be observed in comparatively low magnetic fields $B<1T$.
For this reason the above pointed
$Bi_{1.7}Pb_{0.3}Sr_{2}Ca_{2}Cu_{3}O_{y}$ system was chosen.

If the cylinder-shaped high-temperature layered superconductor sample being
in the mixed state is suspended by a thin elastic thread and set
in axial-and-torsional vibrations in the transverse magnetic field,
then vortexes formed in the sample and tending to be oriented
along magnetic field should increase the elastic moment acting
from a suspend on the sample because they are fixed at the pinning
centers \cite{andro68}. On the other hand, vortexes being torn from the pinning
centers at some torsional angles should increase the dissipation
of the sample at their motion relative to it. Figure 1 gives a
diagram of the instrument. Sample 1 was glued to the crystal
holder 2, and attached to straightened glass rod 4, which, in its
turn, was suspended on the thin elastic thread 3. In the upper part
of the rod, light organic glass disk 5 of diameter R=1 cm and mass
m=0.4g was fixed.

The moment of inertia of the suspend system was $I=0.5g\cdot
cm^{2}$. The recording of vibrations was made using mirror 6 glued
to the suspend system, and photoresistive transducers, mounted  on
scale 7. The scale with transducers was separated from mirror 6 at a
distance of l=200cm and the maximal vibration amplitude in our
experiments didn't exceed $\varphi\approx3\cdot 10^{-1}$  rad.

The logarithmic damping decrement $\delta$ of the vibrations of a
suspended system was measured by recording the time of flight
of a light spot, reflected from the vibrating mirror, between
two transducers\cite{andro59}. The part of cryostat, where a superconductive
sample was placed, was filled with liquid helium or liquid
nitrogen, and placed in a 8kOe magnetic field perpendicular to the
superconductive cylindrical sample axis. Electromagnet 8 was
supported by a system of bearings and could rotate around its
axis. The system of coils 9, 10, 11 can provide both
longitudinal and transverse fields in respect to sample's axis. In
the case of necessity these coils also could be useful for
application to the sample of additional pulsed and oscillating
magnetic fields. The sample was glued to the crystal holder in such a
way that its c-axis was directed along the magnetic field H
perpendicular to the vibration system axis. By rotating of
electromagnet 8 one could set the direction of magnetic field
along c-axis using the diamagnetic moment of forces acting on the
c-oriented sample \cite{ash01}.

If necessary, we could also measure the anisotropy of
elastic-viscous properties of vortex system.

The small axial vibrations of sample were measured during which
its {\bf c}-axis experienced small deviations with respect to the
magnetic field. The maximal amplitude of vibrations was not more
than $\varphi_{max}\approx3\cdot 10^{-1}$ rad.  The dependence of
logarithmic damping decrement of the vibrations on the outer
magnetic field tension {\bf H} directed along {\bf c}-axis of
strongly anisotropic layered high-temperature superconductor
system $Bi_{1.7}Pb_{0.3}Sr_{2}Ca_{2}Cu_{3}O_{y}$  was
investigated.

In Fig.2 the dependence of d on the outer magnetic field tension H
is presented. As it is seen from the figure, in the weak fields up
to approximately H=140 Oe the damping $\delta$ with an increase of the
magnetic field tension increases, but then beginning from 175 Oe
it sharply decreases and already at magnetic fields of the order
of 1250 Oe it becomes equal to the damping value at the absence of
a magnetic field.

Further a slow reduction follows, and at field H=1750 Oe and
higher it turns out to be slightly less than at the absence of
magnetic field.

A sharp reduction of damping processes at fields $>$175 Oe, to our
opinion , is related with the 3D-2D phase transition in the vortex
system, when volume 3D vortexes transform in quasi-two-dimensional
2D vortexes and the increase of critical current   connected with
this transition and, correspondingly, of the pinning force,
strongly reduces the dissipation processes, connected with the
motion of vortex system. The 175-1750 Oe interval is a transition
range. The possibility of 3D-2D transition was discussed in the very
interesting work \cite{yeh94} in connection with the transformation of V-I
characteristics of $Tl_{2}Ba_{2}Ca_{2}Cu_{3}O_{10-\delta}$ in the
magnetic field range 1500-1600 Oe . The estimation of 3D-2D
transition critical field value $B_{2D}$ according to formula (1)
and using the data of \cite{vin90,sang91}: $\Gamma=3\cdot 10^{3}$, and d=18.5
\AA, gives $B_{2D}\approx2000$ Oe being in our opinion in a good
agreement with the experiment. The estimation of the crossover
temperature $T_{cr}$ by formula (2) using the results of works
\cite{vin90,sang91,lan94}: $T_{c}=107K$, $\xi_{c}(0)=1.4$\AA, and d=18.5\AA
 $ $ results in $T_{cr}$=105.7K.

Beginning from T=90K up to T=105K the $\delta(H)$  dependence
character presented in Fig.2 is not changed. But at T=106K the
$\delta(H)$ value is sharply reduced and basically changed that
points to the correctness of our estimation of $T_{cr}$.

In our opinion, the transition range presented in the increased
scale in the insertion to Fig.2 deserves an attention.

As it is seen in Fig.2 at the weak fields $H<1$ kOe the
$\delta(H)$ dependence shows the presence of peaks which can be
connected with phase transitions in the vortex system, for example,
the transition of Abrikosov vortex \cite{abr57} into glassy state or its
melting.

The estimation of vortex lattice melting field \cite{bla} gives:
\begin{equation}
B_m (T) \approx \beta _m \frac{{C_L ^4 }}{{G_i }}H_{c2} (0)\left(
{1 - \frac{T}{{T_c }}} \right)^2
\end{equation}
where $\beta_{m}\approx5$, $C_{L}\approx0.2$
\begin{equation} G_i
= \frac{{T_c }}{{\sqrt 2 \varepsilon _0 (0)d}}
\end{equation}
\begin{equation}
\varepsilon _0 (0) = \left( {\frac{{\Phi _0 }}{{4\pi \lambda (0)}}} \right)^2
\end{equation}
The second critical field $H_{c2}$ is estimated in the following
way\cite{shm82}:
\begin{equation}
H_{c2} (0) = \frac{{\Phi _0 }}{{2\pi \xi _{ab}^2 (0)}}
\end{equation}
where $\xi_{ab}(0)$ in different experiments falls in the range
$15\div17$ \AA.

One obtains: $H_{c2}(0)=(114\div146,5)$ T. Hence, using formulas
(3),(4) and (5) one obtains: $B_{m}(T)\approx275\div315$ Oe at
T=100 K.

This value is in a good agreement with the second maximum of
logarithmic damping decrement of vibrations in the insertion of
Fig.2 which is seen at $H\approx280$ Oe.

So, the damping peak observed by us in the transition range of
magnetic fields can be related with the Abrikosov vortex melting,
i.e. with the melting of 3D vortexes\cite{bla}.

It should be pointed that the 3D-2D transition at increasing
magnetic field, predicted by M.Fisher and collaborators \cite{fish91} in
the strongly anisotropic layered materials, can be realized also in
superconductors with not so extremely high value of anisotropy
factor, as in BSCCO, but with its comparatively low value.

As an illustration one can consider the interesting work \cite{wol94} on
investigation of the 3D-2D transition in the fine layers of
$YBa_{2}Cu_{3}O_{7-\delta}$.

In conclusion, it should be pointed that the sharp decay of
dissipation at fields B$>$Â175 Oe observed by us is probably
connected with a sharp increase of pinning force because the
amplitude effects of dissipative processes sharply decrease
at strong fields B$>$Â2000 Oe.

Thus, this phenomenon, in its turn, points again to the
possible phase transition in the 3D-2D vortex structure in fields
B$>$Â2000 Oe.

We have started experiments on the direct measurement of pinning
force at the 3D-2D phase transition.

And, finally, authors are expressing their sincere gratitude to
M. J. Chubabria for preparing the samples, and to N. G. Suramlishvili
and Yu. G. Mamaladze for the discussion of
experimental results.

This work is made with support of International Scientific and
Technology Center (ISTC) through Grant  G-389.


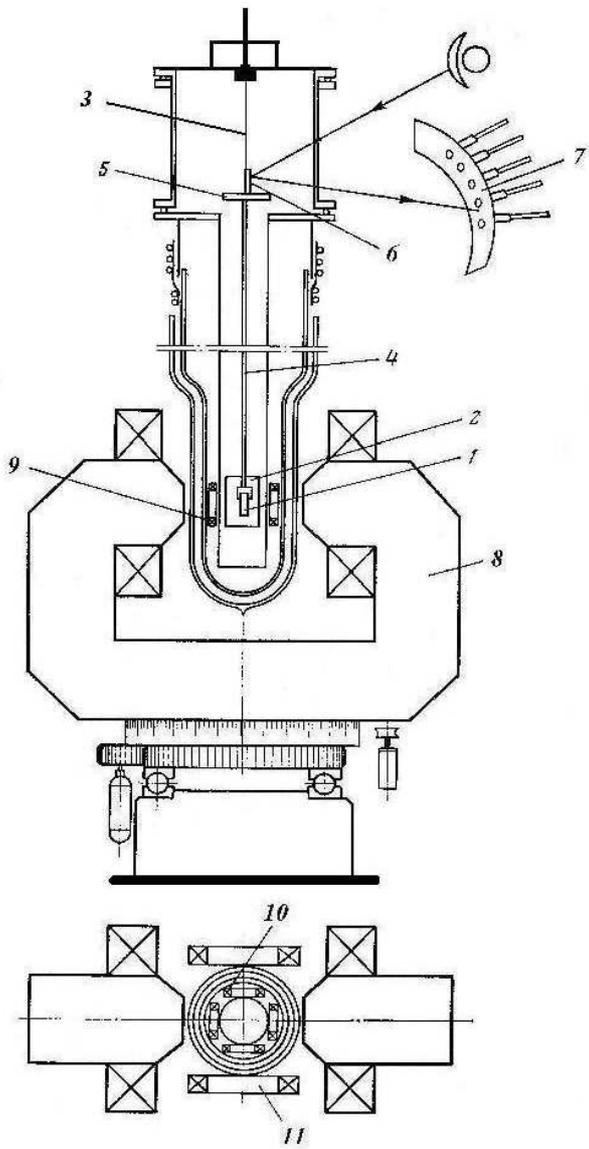
\begin{figure}
\caption{ The diagram of low-temperature part of instrument:
1 - sample, 2 - crystal holder, 3 - thread, 4 - glass rod , 5 - discus, 6 - mirror,
7 - scale with photoresistive transducers, 8 - electromagnet, 9-11 - coils.
}
\label{fig1}
\end{figure}

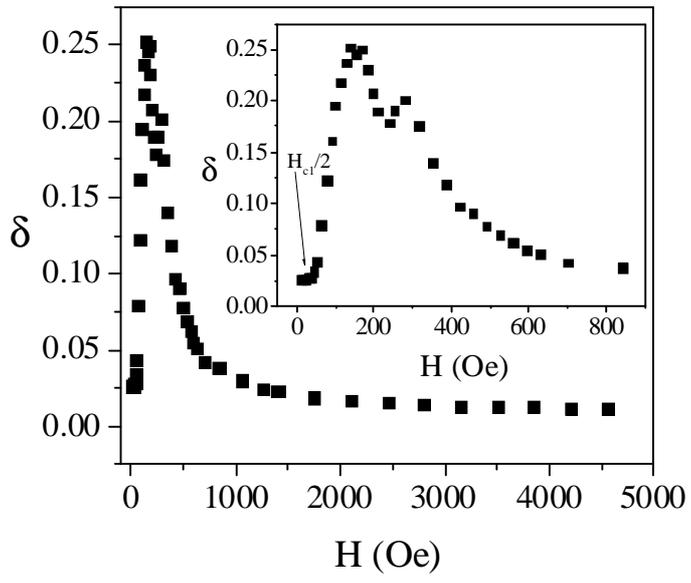
\begin{figure}
\caption{Logarithmic damping decrement $\delta$ of the vibrations
of a suspended system as a function of the external magnetic field
H.} \label{fig2}
\end{figure}


\begin{references}

\bibitem{jab73}
J. G. Chigvinadze,
Zh. Eksp. Teor. Fiz. {\bf 65}, 1923 (1973).

\bibitem{ols00}
C. J. Olson, G. T. Zimanyi, A. B. Kolton and N. Gronbech-Jensen,
Phys.Rev. Lett. {\bf85}, 5416 (2000);
C. J. Olson, C. Reichbardt, R. T. Scalettar and G. T. Zimanyi,
 arXiv:cond-mat/0008350 v.1  23 Aug 2000, 1-4.

\bibitem{feig90}
M. V. Feigelman, V. B. Geshkenbein and A. I. Larkin, Physica
C {\bf 167}, 177 (1990);

\bibitem{vin90}
V. M. Vinokur, P. H. Kes and A. E. Koshelev
Physica C {\bf168}, 29 (1990).

\bibitem{far89}
D. E. Farrell, S. Bonham, J. Foster, Y. C. Chang, P. Z. Jiang,
K. G. Vandervoort, D. L. Lam and V. G. Kogan,
Phys. Rev. Lett. {\bf63},782(1989).

\bibitem{andro68}
E. L. Andronikashvili, S. M. Ashimov, J. S. Tsakadze, J. G. Chigvinadze,
Zh. Eksp. Teor. Fiz. {\bf55},775 (1968).

\bibitem{andro59}
E. L. Andronikashvili, Yu. G. Mamaladze, J. S. Tsakadze,
Trudi Inst. Fiz. ANGSSR, 7 (1959);
J. G. Chigvinadze, A. G. Djagarov, V. S. Nadareishvili, T. A. Japiashvili,
Prib. Tekh. Eksp. {\bf192} (1984).

\bibitem{ash01}
S. M. Ashimov, J. G. Chigvinadze.
Prib. Tekh. Eksp. 2001 (in press).

\bibitem{yeh94}
W. J. Yeh, L. K. Yu, Z. Q. Yu, Y. Xin and W. K. Wong,
Physica B, {\bf194-196}, 1485 (1994).

\bibitem{abr57}
A. A. Abrikosov,
Zh. Eksp. Teor. Fiz. {\bf32}, 1442 (1957).

\bibitem{sang91}
Sang-Geun Lee,  Young-cheol Kim,  Chong-You Park, Seong-cho Yu and Min-Su Jang
Physica B, {\bf169}, 657 (1991).

\bibitem{lan94}
W. Langa, W. Kula, R. Sobolevski,
Physica B, {\bf194-196}, 1643 (1994).

\bibitem{bla}
G. Blatter, M. V. Feigelman, V. B. Geshkenbein, A. I. Larkin and V. M. Vinokur,
Teoretishe Physek, EidgenossisheTechnishe Hochschule Zurich-Hongerberg
CH=8093 (Zurich, Switzerland).

\bibitem{shm82}
V. V. Shmidt,
{\it Introduction to Superconductor Physics},
(Nauka, Moscow, 1982) p. 133.

\bibitem{fish91}
M. P. A. Fisher {\it et al}., Physica B {\bf169}, 85(1991).

\bibitem{wol94}
 P. J. M. Woltgens, C. Dekker, R. H. Koch, B. W. Hussly
and A. Gupta,
Physica B {\bf194-196}, 1911 (1994).


\end{references}
\end{document}